\documentclass[10pt, conference]{IEEEtran}

\usepackage{enumerate}
\usepackage[utf8]{inputenc}
\usepackage{subcaption}
\usepackage{epsfig}
\usepackage{caption}
\usepackage{etoolbox}
\usepackage{wrapfig}
\usepackage{colortbl}
\usepackage{graphicx}

\usepackage{balance}
\usepackage{cite}
\usepackage[numbers,sort&compress]{natbib}
\usepackage{epsfig}
\usepackage{times}
\usepackage{graphicx}
\usepackage[figuresright]{rotating}
\usepackage{amssymb}
\usepackage{amsmath}
\usepackage{setspace}
\usepackage{rotating}
\usepackage{multirow}
\usepackage{wrapfig}
\usepackage{booktabs}
\usepackage{array}
\usepackage{comment}
\usepackage{pict2e}
\usepackage{epstopdf}
\usepackage{multirow}
\usepackage{hhline}
\usepackage{subcaption}
\usepackage[font=footnotesize]{caption} %
\usepackage[]{nohyperref}  %
\usepackage{url}  %

\usepackage{algorithmic}
\usepackage{algorithm}
\floatname{algorithm}{Algorithm} \textfloatsep 0.5cm

\usepackage{colortbl}

\newcommand{\be}{\begin{equation}}
\newcommand{\ee}{\end{equation}}

\usepackage[bottom]{footmisc}

\begin{document}
\title{Towards Ultra-low-power Realization of Analog Joint Source-Channel Coding using MOSFETs}

\author{{\bf Vidyasagar Sadhu, Sanjana Devaraj, and Dario Pompili}\\
Department of Electrical and Computer Engineering, Rutgers University--New Brunswick, NJ, USA\\
E-mails: vidyasagar.sadhu@rutgers.edu, sd1049@scarletmail.rutgers.edu, pompili@rutgers.edu\\
}

\maketitle

\thispagestyle{empty}
\pagestyle{plain}
\pagenumbering{gobble}

\begin{abstract}
Certain sensing applications such as Internet of Things~(IoTs), where the sensing phenomenon may change rapidly in both time and space, requires sensors that consume ultra-low power (so that they do not need to be put to sleep leading to loss of temporal and spatial resolution) and have low costs (for high density deployment). A novel encoding based on Metal Oxide Semiconductor Field Effect Transistors~(MOSFETs) is proposed to realize Analog Joint Source Channel Coding~(AJSCC), a low-complexity technique to compress two (or more) signals into one with controlled distortion. In AJSCC, the y-axis is quantized while the x-axis is continuously captured. A power-efficient design to support multiple quantization levels is presented so that the digital receiver can decide the optimum quantization and the analog transmitter circuit is able to realize that. The approach is verified via Spice and MATLAB simulations.
\end{abstract}

\section{Introduction}\label{sec:intro}
\textbf{Motivation:}
Wireless Sensor Networks~(WSNs) are currently used for several purposes~\cite{Zhao2017mass, Zhao2018tbcas} including environmental monitoring~\cite{joe17}, infrastructure surveillance~\cite{Kumar16}, and intelligent transportation systems~\cite{Hu15}. Sensors in these networks should be able to capture high spatial and temporal resolution exhibited by the corresponding phenomenon. 
In order to achieve this, Sadhu et al.~\cite{Sadhu2017wons}
have previously proposed a three-tier sensor network architecture consisting of low-power, low-complexity all-analog sensors at Tier~1 that sits below the traditional WSN consisting of digital Cluster Heads~(CHs) at Tier~2 and a fusion center/mobile sink at Tier~3. They call analog sensors that comprise Tier~1 as ``dumb'' sensors as they only sense and transmit but do not perform any processing unlike traditional digital sensing motes. The processing in turn is outsourced to powerful Tier-2 digital CHs, where one CH processes data from thousands of Tier-1 sensors. Tier-1 analog sensors have a low-power and low-complexity design that allows them to sense continuously, and also be deployable in high density thanks to their low cost.
To realize this low-complexity design for Tier-1 sensors, the authors in~\cite{Sadhu2017wons} have proposed to use a low-complexity encoding technique called Joint Source-Channel Coding~(JSCC), also known as Shannon mapping~\cite{Shannon49}, which compresses two (or more) signals into one. 
JSCC achieves this using a space-filling curve where the x-axis signal is continuously captured while the y-axis one is quantized (let us denote the amount of quantization as $\phi$). The sensed $(x,y)$ point is mapped to the closest point on the curve and the encoded (compressed) value is a property of the curve, e.g., length of the curve from origin. By adopting this technique for Tier-1 analog sensors, Sadhu et al. have argued that their architecture is a suitable candidate for high-bandwidth applications such as Internet of Things~(IoT) due to the high spectral efficiency achieved via compression~\cite{Sadhu2017wons}. %
To achieve the low-power/low-complexity advantages of JSCC, this technique needs to be realized in the analog domain---hence the name Analog JSCC or AJSCC. However, AJSCC is hard to realize on hardware in a power-efficient manner. %
This is especially important when the sensors are powered using energy-harvesting techniques~\cite{IoUT}. %

\textbf{Our Vision:}
To realize JSCC in an energy-efficient manner, we take a completely different path compared to previous approaches that implement rectangular Shannon Mapping~\cite{Zhao16}. We propose to realize JSCC using the input-output (also called IV, which stands for current-voltage) characteristics of a single Metal Oxide Semiconductor Field Effect Transistor~(MOSFET) device as the space-filling curve for JSCC.
Through this novel approach, we are able to achieve power of the order of few tens of $\mu \rm{W}$ (possibility of few $\mu \rm{W}$, as explained in Sect.~\ref{sec:perf_eval}) for the encoding circuit compared to several tens of $\mu \rm{W}$ in previous circuits. We are able to achieve this by working fully in the analog domain and by avoiding power-hungry Analog to Digital Converters~(ADCs) and microprocessors, which are used in digital sensing motes. 

\textbf{Related Work:}
Most of the existing JSCC-hardware solutions are all digital and power hungry. For example, a Software-Defined Radio~(SDR) system to realize AJSCC mapping has been reported in~\cite{Garcia11}. The mapping was also recently implemented in an optical digital communication system in~\cite{Romero14} and has been combined with Compressive Sensing~(CS) in~\cite{Saleh12} to improve robustness against channel noise. Shannon mapping encoding was adopted in~\cite{stopler14} for a digital video transmission. All these design solutions use digital microcontrollers, which are quite power hungry: for example, in~\cite{TI11}, with a $1.8~\rm{V}$ supply, the power consumption of a microcontroller alone can be as high as $450~\rm{mW}$ ($250~\rm{mA} \times 1.8~\rm{V}$). 
However, there are a couple of works that try to realize JSCC in analog domain like we do. However, these realize the rectangular JSCC unlike ours where we use a novel space-filling curve. To the best of the authors' knowledge, ours is the first work to realize a different space-filling curve in hardware analog domain than the rectangular JSCC. Among those that realize rectangular AJSCC, Zhao et al.~\cite{Zhao16} proposed all-analog sensor design that realizes AJSCC using Voltage Controlled Voltage Sources~(VCVSs). This design, which they call, ``Design~1'' is an inefficient design as it adopts a fixed number of JSCC levels, hardware in each stage is duplicated, and does not scale with the number of levels. 
Even though the authors proposed ``Design~2''~\cite{Zhao2018} to address the above limitations, it still has higher power consumption. Design~1 with 11 JSCC levels (quantization levels on y-axis) consumes $130~\rm{\mu W}$, whereas Design~2 with 16 levels consumes $72~\rm{\mu W}$ ($64~\rm{\mu W}$ for 8 levels). These numbers, which do not include the transmission power, are large for sensors powered using energy-harvesting techniques that produce only tens of $\mu \rm{W}$~\cite{IoUT,Khan2018} to power the entire sensing/transmitting device. Transmission power is around $50~\mu\rm{W}$ using MEMS-based transceivers with a range of few meters~\cite{Kim2015, memstransceiver1}. Differently from above, we adopt the MOSFET's IV characteristics as the space-filling curve and are able to achieve encoding power consumption of $\approx 24~\mu\rm{W}$, with possibility of $8~\mu\rm{W}$.

\textbf{Our Contributions} can be summarized as---(i)~we propose a MOSFET-based encoding method to realize AJSCC (with MOSFET input-output characteristics as the space-filling curve) that consumes power of only few tens of $\rm{\mu W}$; (ii)~we propose a circuit design to accommodate different levels of quantization in the y-axis ($\phi$) of JSCC; (iii)~we verify the functionality of the approach by studying the receiver via both Spice and MATLAB simulations.

\textbf{Paper Outline:}
We present our approach in Sect.~\ref{sec:prop_soln},
evaluate it via simulations in Sect.~\ref{sec:perf_eval}, and
finally conclude and provide future research directions in Sect.~\ref{sec:conc}.

\section{Proposed Solution}\label{sec:prop_soln}
We first present our novel idea of using MOSFET to realize AJSCC along with its associated challenges, and then describe our precircuit, which allows for different values for $\phi$.

\begin{figure}
\begin{center}
\includegraphics[width=3.6in]{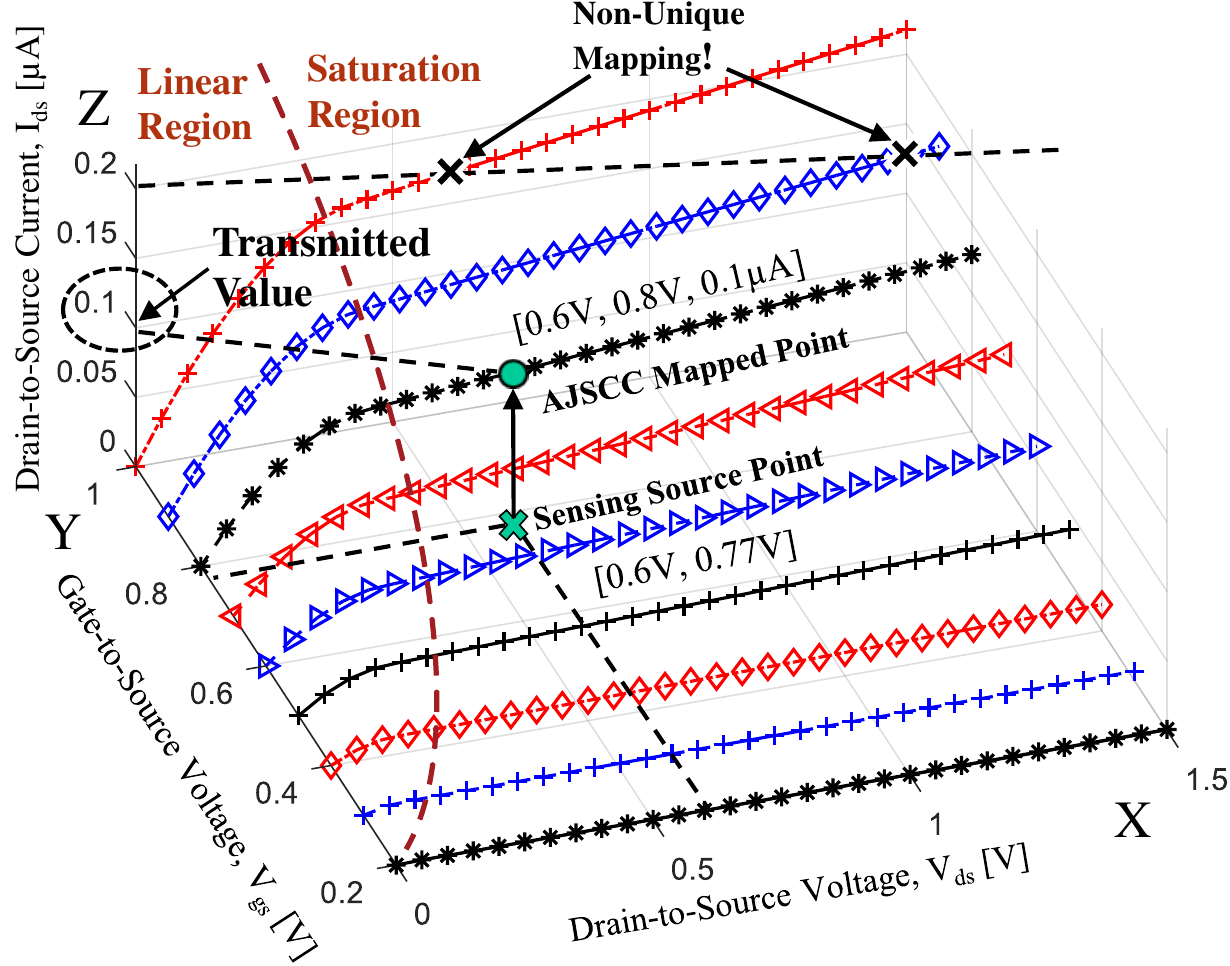}
\end{center}
\vspace{-0.15in}
\caption{Shannon mapping realized via output characteristics ($I_{ds}$ vs. $V_{ds}$ for different $V_{gs}$) of a MOSFET in saturation region (right of dashed line).}\label{fig:ajscc_mos_3d}
\vspace{-0.2in}
\end{figure}

\textbf{MOSFET-based Encoding at Transmitter:}
\begin{figure*}[ht]
        \centering
           \begin{subfigure}[b]{0.3\textwidth}
        		\centering
        		 \includegraphics[width=1\textwidth,height=0.9in]{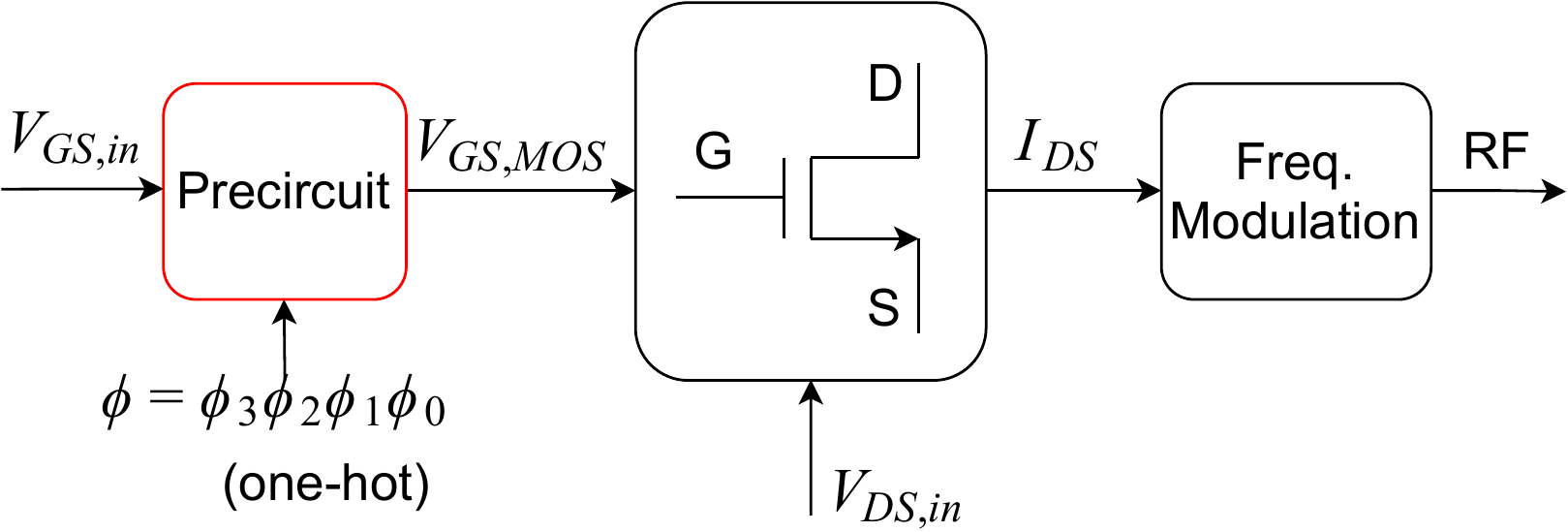}
        		\caption{}
        		\label{fig:mos_encoding}
        	\end{subfigure}
        	\hspace{-0.1in}
~
        \begin{subfigure}[b]{0.68\textwidth}
            \centering
            \includegraphics[width=1.08\textwidth,height=1in]{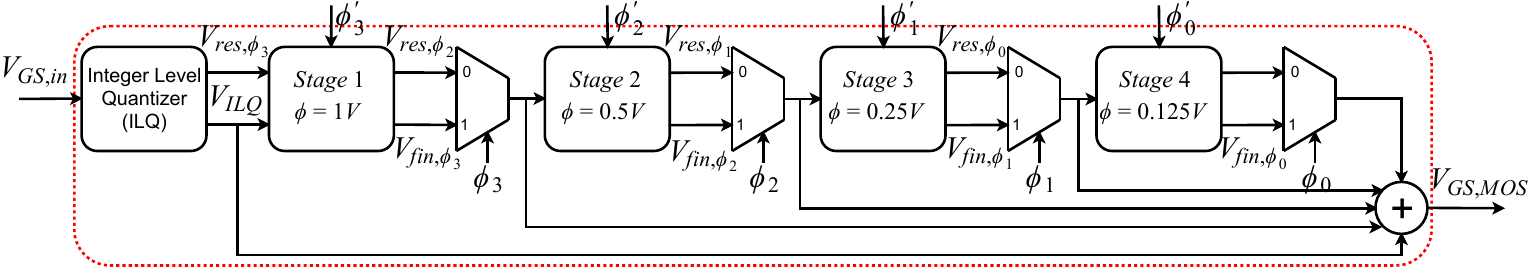}
            \caption{}
            \label{fig:precircuit}
        \end{subfigure}
        \vspace{-0.1in}
        \caption{\label{fig:mos} (a)~MOSFET-based encoding for Analog Source-Channel Coding~(AJSCC) with different levels of $\phi$; (b)~Block diagram of precircuit used in (a).}
        \vspace{-0.1in}
\end{figure*}
\begin{figure*}
\begin{center}
\includegraphics[width=7.1in]{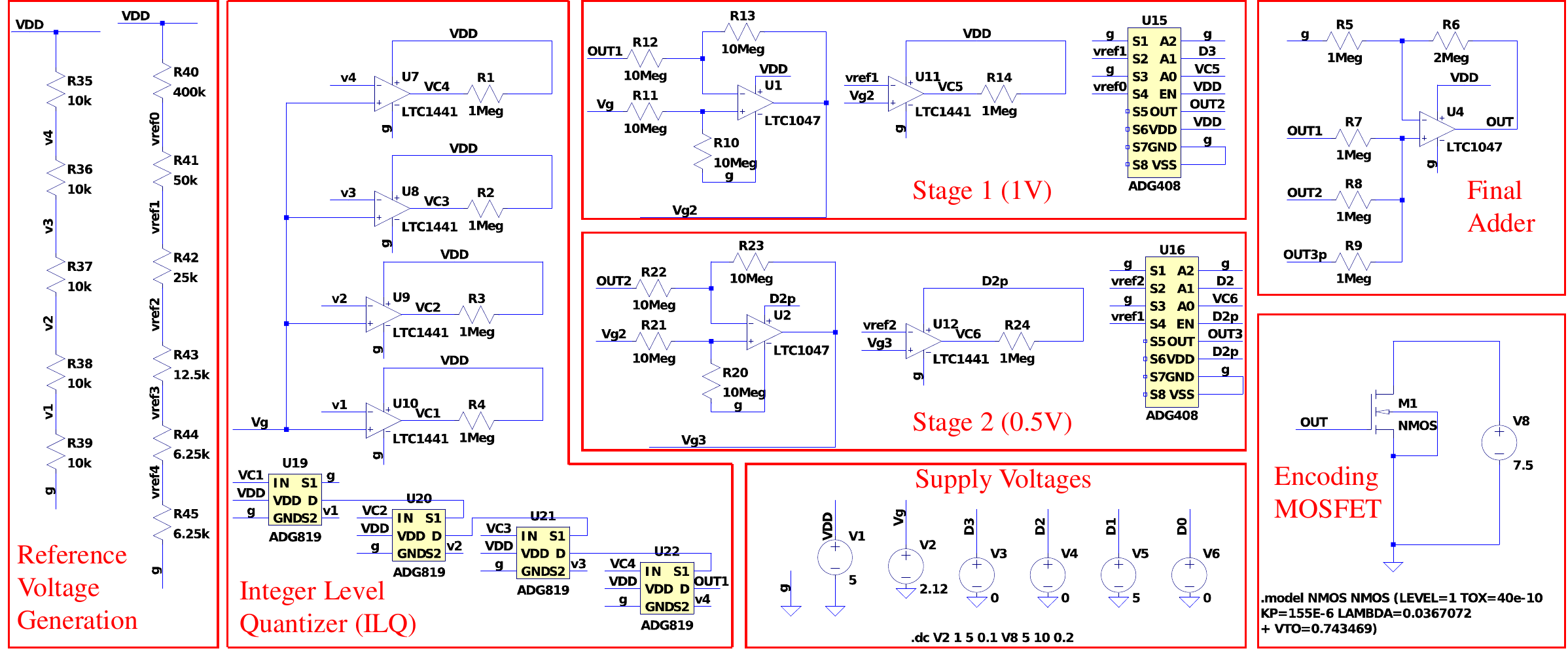}
\end{center}
\vspace{-0.1in}
\caption{Spice implementation of precircuit and MOSFET-based encoding in Fig.~\ref{fig:mos_encoding}. Other stages and simple logic generating $\phi'$ are not shown.}\label{fig:circuit_paper}
\vspace{-0.25in}
\end{figure*}
Ideally, any new space-filling curves for AJSCC should preserve these properties: ($i$)~they should achieve better trade-off between channel noise/compression and approximation noise; ($ii$)~they should be realizable using \emph{all-analog} components; and ($iii$)~they should result in a \emph{unique} mapping (i.e., two or more sensor values should map to \textit{only one} AJSCC encoded value). Given these desirable properties of a space filling curve, we propose the idea of using the IV charactersitics of a MOSFET in saturation region as the space-filling curve (instead of using rectangular parallel lines as used in~\cite{Zhao16, Zhao2018}). 
A MOSFET has three terminals: Gate~(G), Drain~(D), and Source~(S). When a suitable voltage is applied across G and S terminals, $V_{gs}$, and D and S terminals, $V_{ds}$, a current is generated across D and S terminals, $I_{ds}$.
The relationship among $V_{gs}$, $V_{ds}$, and $I_{ds}$ for a \emph{real} MOSFET in the saturation region (Fig.~\ref{fig:ajscc_mos_3d}) is, 
\begin{equation} \label{eq:ids_clm}
I_{ds}=\frac{1}{2} \cdot \frac{W}{L} \cdot {\mu}C_{ox} \cdot (V_{gs}-V_{th})^2 \cdot (1+{\lambda} V_{ds}),
\end{equation} where $W,L~[\rm{m}]$ are width and length of the MOSFET channel, respectively, $\mu~[\rm{m^2/Vs}]$ is the electron mobility in the channel, $C_{ox}~[\rm{F/m^2}]$ is the oxide capacitance per unit area, and $\lambda~[\rm{V^{-1}}]$ is the Channel Length Modulation~(CLM) parameter. Because of CLM, $I_{ds}$ keeps increasing at a very slow rate (governed by $V_{gs}$ and other parameters) in the saturation region.

Fig.~\ref{fig:ajscc_mos_3d} shows these $I_{ds}$ curves in the saturation region to the right of dashed line, generated via Spice, where
$V_{gs}$ is varied in the discrete set, $0.2,0.3,...,1~\rm{V}$ ($28~\rm{nm}$ Silicon technology model MOSFET is used for illustration purpose).
We can notice that the slope of the current curves increases as $V_{gs}$ increases due to CLM, which we leverage to perform the decoding at the receiver, as explained below.
$I_{ds}$ encodes the values of $V_{gs}$ and $V_{ds}$ (as opposed to extracting the length of the curve from origin to the mapped point, as in~\cite{Zhao16, Zhao2018}). 
It is necessary to have a discrete set of y-axis ($V_{gs}$) values, and the actual y-axis value is mapped to the nearest value from the set and applied to the MOSFET to generate the encoded current (Fig.~\ref{fig:ajscc_mos_3d}).
While the proposed MOSFET-based space-filling technique satisfies ($i$) and ($ii$) properties mentioned above, it violates ($iii$) as a given $I_{ds}$ value
could be generated from multiple pairs of $V_{gs}$ and $V_{ds}$ values (Fig.~\ref{fig:ajscc_mos_3d}).
This is problematic as it is difficult to decode the correct $V_{gs}$ at the receiver. To address this challenge, we propose a decoding technique at the digital receiver based on the previously received $I_{ds}$ value.

\textbf{Decoding at Receiver:}
We assume that the discrete set of $V_{gs}$ values 
used at the analog transmitter for encoding is known at the digital receiver. This is a valid assumption as the receiver decides the optimum $\phi$ to be used by the transmitter~\cite{Zhao2018}. The decoding process relies on the assumption that physical values do not change abruptly and hence two consecutive received $I_{ds}$ values at the receiver will lie on the same $I_{ds}$ curve (i.e., corresponding to a particular $V_{gs}$ value). The probability of them lying on different $I_{ds}$ curves (i.e., corresponding to different $V_{gs}$ values) is low as the two consecutive values would have sensed similar values (i.e., the sampling rate at the sensor is more than the rate of change of the phenomenon), would have experienced similar wireless channel conditions (i.e., the sampling rate at the sensor is more than the rate of change of channel) and hence would belong to the same $I_{ds}$ curve. The challenge then lies in identifying the correct $V_{gs}$ value out of the discrete set of $V_{gs}$ values used at the transmitter using these two consecutive $I_{ds}$ values. For this purpose, we make use of a \emph{slope-matching technique}~\cite{Sadhu2018ucomms}---we pick that $V_{gs}$ curve whose slope matches closely with the slope calculated theoretically using~\eqref{eq:ids_clm} $\approx \lambda \cdot I_{ds}$.

\textbf{Variable $\phi$ Design:}
There are scenarios in which there is a need to change the $\phi$ \emph{adaptively}, e.g., temperature in one of the north-eastern states of United States will likely be in the range of -$10$ to $35^\circ$C. As such we need high resolution within this range and low-accuracy outside it. Also, within this range, the temperature will have a certain distribution; so, it is desirable to have a varying accuracy (and, so, variable $\phi$) over the entire range.
Hence, we enable our design to accept different levels of $\phi$, specifically, we design for $\phi=1, 0.5, 0.25, 0.125~\rm{V}$. 

Fig.~\ref{fig:mos_encoding} shows the high-level design of our circuit, which consists of a precircuit followed by MOSFET-based encoding. The function of the precircuit is to take the raw voltage ($V_{gs,in}$) from the sensor and then quantize it as per $\phi$ to obtain the $V_{gs}$ voltage to be fed to MOSFET ($V_{gs,MOS}$). There is no need of any processing for the other sensor voltage ($V_{ds}$). These voltages are then supplied to the MOSFET to generate the encoding current value ($I_{ds}$), which is then frequency modulated and transmitted via Radio Frequency~(RF) (precircuit in Fig.~\ref{fig:mos_encoding} is further expanded in Fig.~\ref{fig:precircuit}). We represent input $\phi$ via four voltages, $\phi_3,\phi_2,\phi_1,\phi_0$, in a one-hot encoding manner; e.g., for $\phi=1~\rm{V}$, we have, $\phi_3=5~\rm{V},\phi_2=\phi_1=\phi_0=0~\rm{V}$ (where $5~\rm{V}$ indicates HIGH; $0~\rm{V}$ LOW). 

The precircuit consists of an Integer-Level Quantizer~(ILQ), which first quantizes the input voltage ($V_{gs,in}$) to its floor value, $\left \lfloor{V_{gs,in}}\right \rfloor$. The successive stages then find the appropriate residual value based on the provided $\phi$. There are four stages corresponding to the four $\phi$ values. The input to stage~1 is the residual difference, $V_{gs,in} - \left \lfloor{V_{gs,in}}\right \rfloor$. Each stage takes the residual voltage from the previous stage as input and generates one of the two voltages, $V_{res}$ and $V_{fin}$, standing for residual (which is given as input to the next stage) and final values. Whenever a particular stage is the last stage for a given $\phi$ e.g., stage~2 for $\phi = 0.5~\rm{V}$, the stage generates the $V_{fin}$ value, else the stage generates $V_{res}$ value. Final and residual values generated by a stage will depend on which stage it is and what the $\phi$ value is. For stage~$n$ ($n=1,2,3,4$), i.e., $\phi=1/2^{n-1}~\rm{V}$, when input residual $<0.5/2^{n-1}~\rm{V}$, output residual $V_{res}=0~\rm{V}$ and $V_{fin}=0~\rm{V}$; when input residual $\geq 0.5/2^{n-1}~\rm{V}$, output residual $V_{res}=0.5/2^{n-1}~\rm{V}$ and $V_{fin}=1/2^{n-1}~\rm{V}$.
We know if the stage is the last stage or not from the one-hot encoding of $\phi$, which is $\phi_3\phi_2\phi_1\phi_0$, e.g., if $\phi_1=5~\rm{V}$, Stage~3 is the last. If a particular stage is the last for a given $\phi$, the succeeding stages are powered down to save power; hence, the output of those stages is $0~\rm{V}$. This is achieved by generating $\phi_3'\phi_2'\phi_1'\phi_0'$ from $\phi_3\phi_2\phi_1\phi_0$ such that whenever a particular $\phi_i$ is $5~\rm{V}$, all the preceding $\phi_i'$ are also made $5~\rm{V}$. These $\phi_i'$ values are provided as supply voltages, (e.g., $\phi_2'=D2p$ for Stage~2 in Fig.~\ref{fig:circuit_paper}) to each stage so that stages next to the last stage are powered down.

Fig.~\ref{fig:circuit_paper} shows the spice implementation of our design in LTSpice with different blocks indicated as per Fig.~\ref{fig:mos_encoding}. Reference integer voltages and residual voltages ($0.5/2^{n-1}~\rm{V}$) are first generated. 
In the ILQ block, we compare the input voltage to reference integer voltages and use a cascade of 2x1 multiplexers to generate the output. In the Stage~1 block, we first subtract the output of ILQ from $V_{gs,in}$ to generate input residual voltage, which is then compared with reference residual voltage of $0.5~\rm{V}$ and then a 4x1 mux (we used only 4 ports of an 8x1 mux) to generate either $V_{res}$ or $V_{fin}$ voltage, as mentioned above. Similar process is repeated in Stage~2. For illustration purposes, we have shown only two stages in Fig.~\ref{fig:circuit_paper}. 
The outputs of all the stages are added to get $V_{gs,MOS}$.

\begin{figure*}[ht]
        \centering
        \hspace{-0.25in}
           \begin{subfigure}[b]{0.32\textwidth}
        		\centering
        	 \includegraphics[width=1.05\textwidth,height=1.7in]{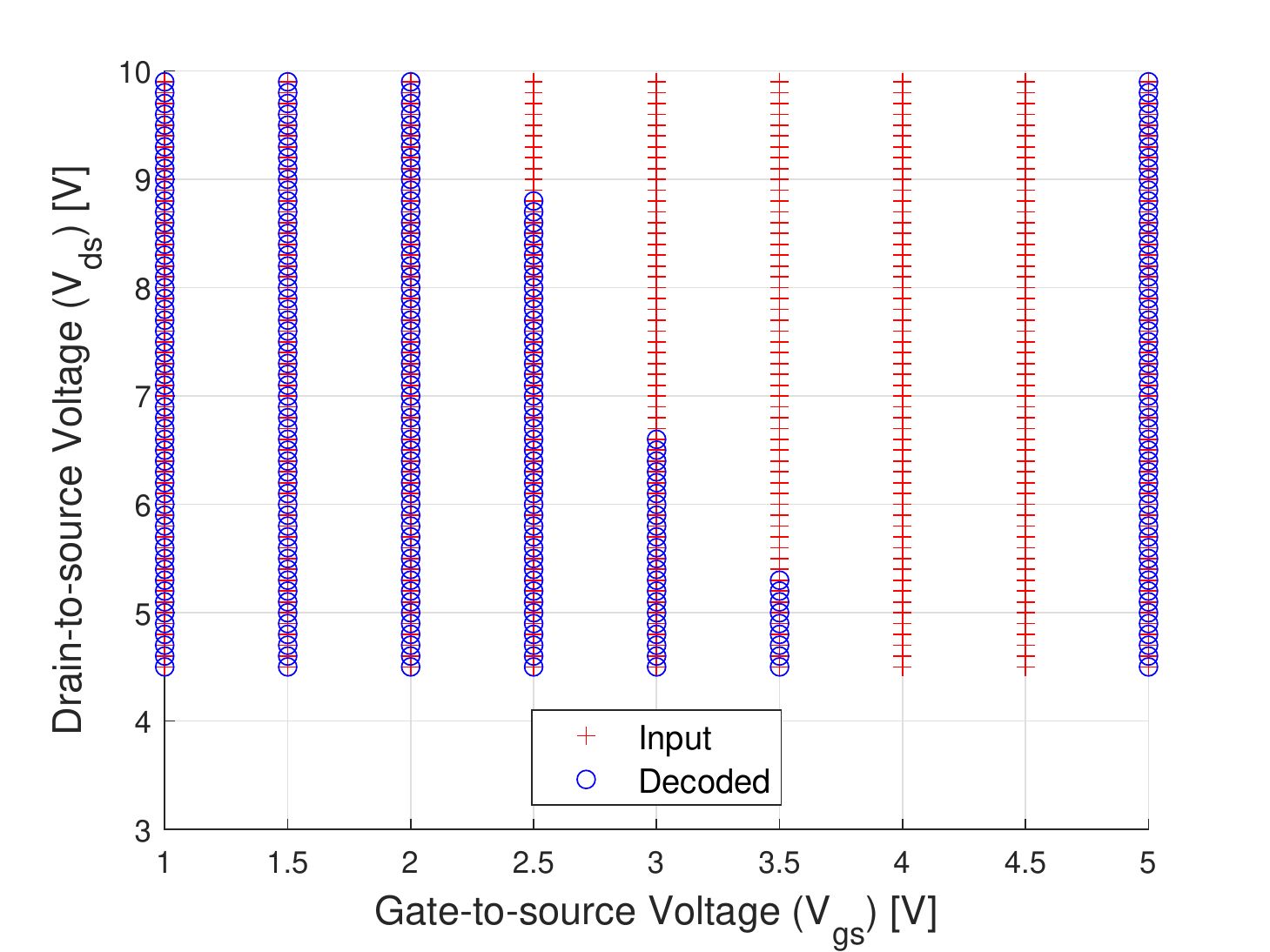}
        		\caption{}
        		\label{fig:before_correction_delta_05}
        	\end{subfigure}
        	 \hspace{-0.15in}
~        	 
        \begin{subfigure}[b]{0.32\textwidth}
            \centering
            \includegraphics[width=1.05\textwidth,height=1.7in]{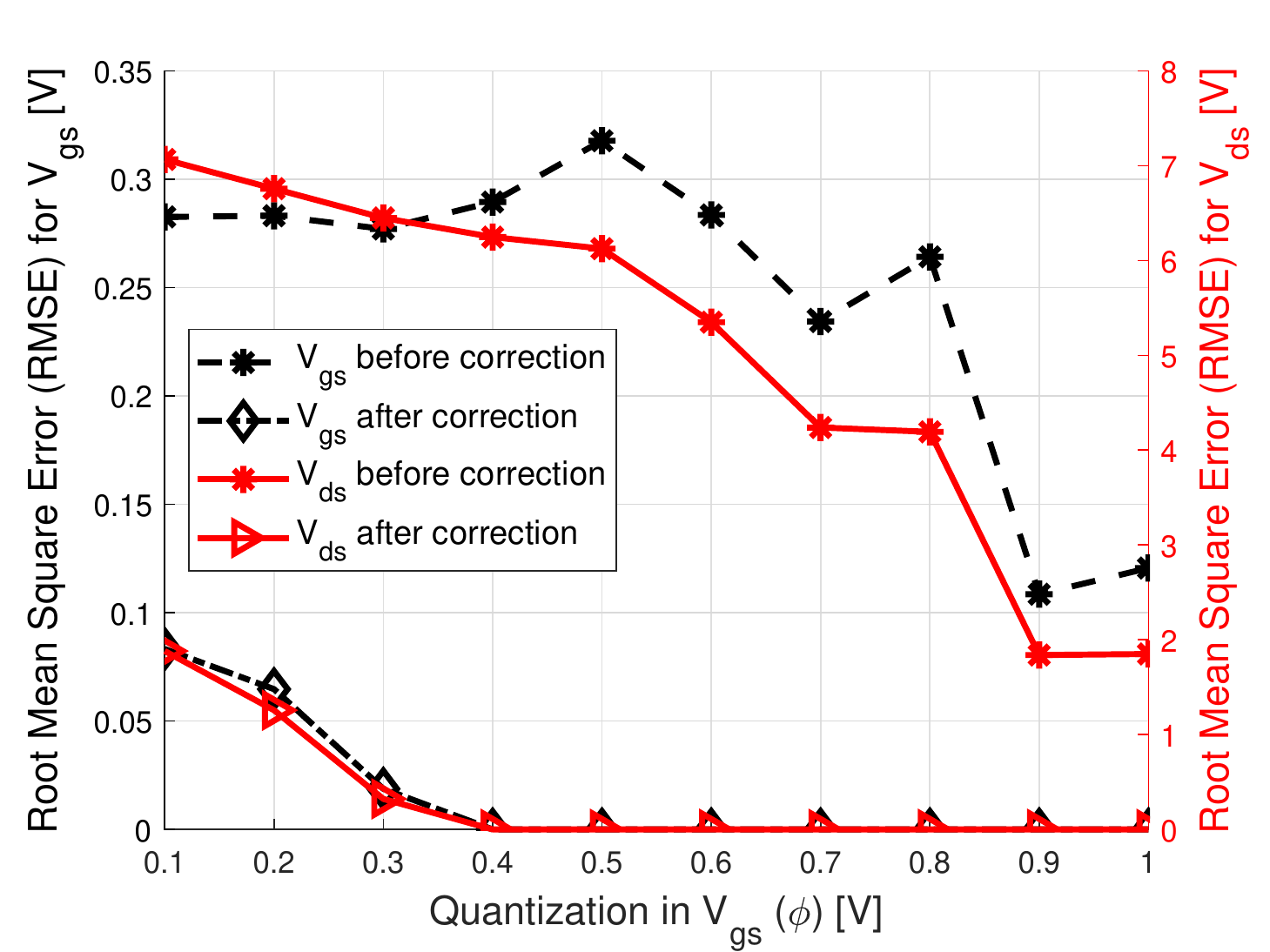}
            \caption{}
            \label{fig:rmse_phi}
        \end{subfigure}
~         
        \begin{subfigure}[b]{0.32\textwidth}
            \centering
            \includegraphics[width=1.09\textwidth,height=1.7in]{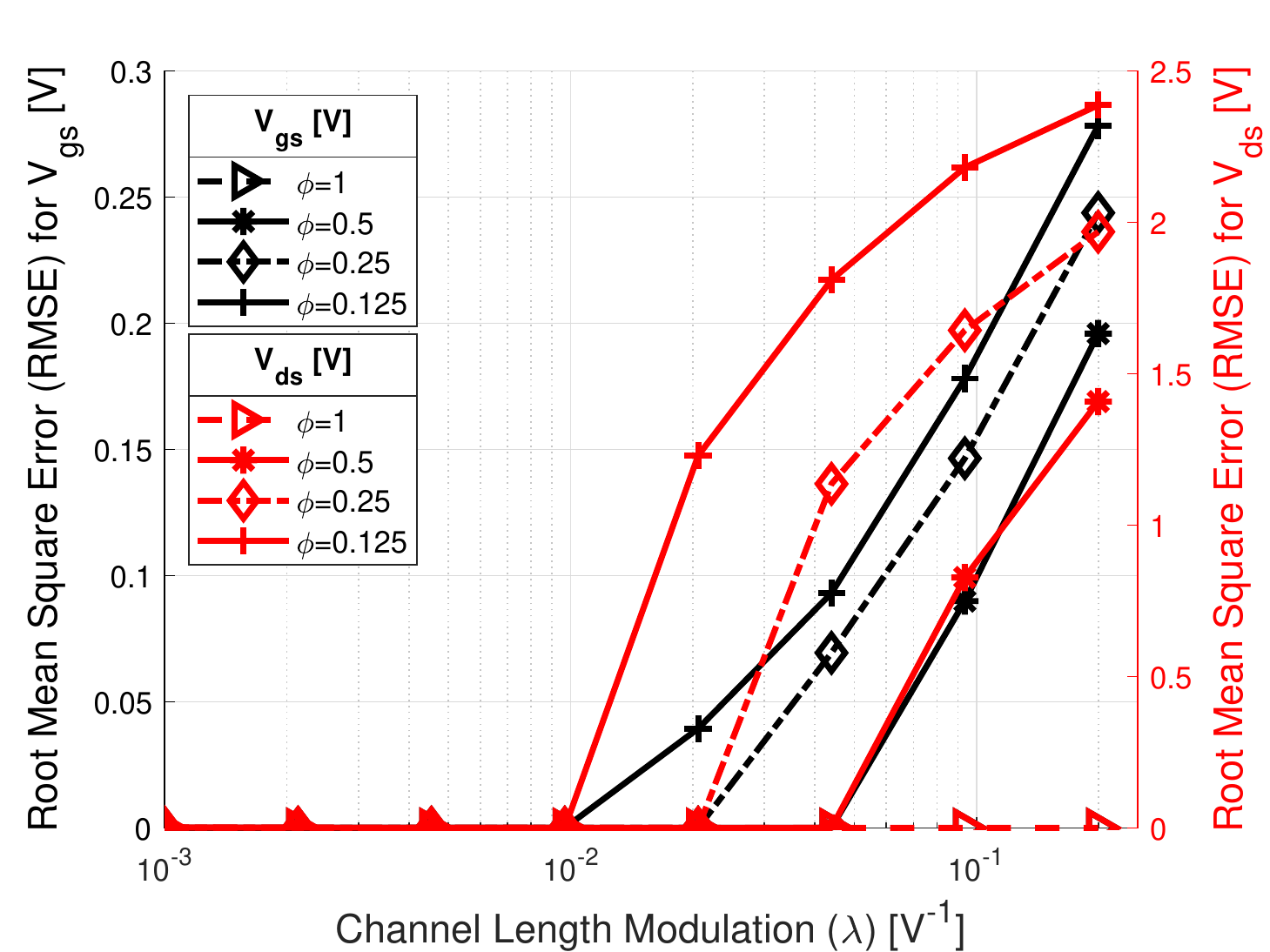}
            \caption{}
            \label{fig:rmse_lambda_phi_vgs_vds}
        \end{subfigure}
        \vspace{-0.1in}
        \caption{\label{fig:precircuit_enc_dec}(a) Decoding results for $\phi=0.5~\rm{V}$ when no correction logic is used; (b)~Root Mean Square Error~(RMSE) of $V_{gs}$ and $V_{ds}$ before and after correction logic is applied as $\phi$ is varied; (c)~RMSE (after the correction is applied) of $V_{gs}$ and ~$V_{ds}$ as $\lambda$ is varied for different values of $\phi$.}
        \vspace{-0.2in}
\end{figure*}

\section{Performance Evaluation}\label{sec:perf_eval}
To verify the functionality of our precircuit as well as of the MOSFET-based encoding and decoding with different levels of $\phi$ we have carried out Spice and MATLAB simulations. 

\begin{figure}
\begin{center}
\includegraphics[width=3in,height=2in]{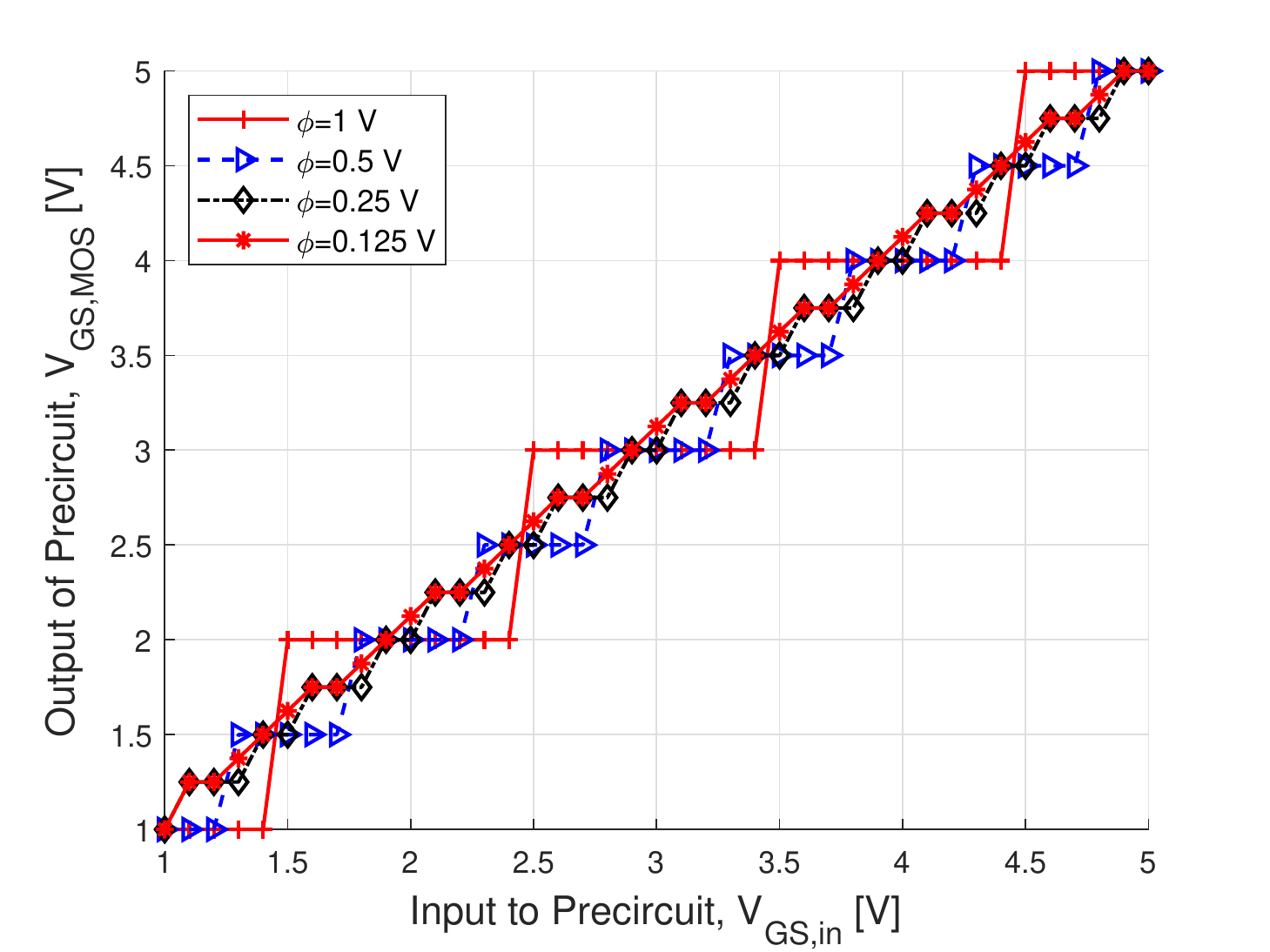}
\end{center}
\vspace{-0.15in}
\caption{~$V_{gs,in}$ mapped to $V_{gs,MOS}$ by the precircuit for different $\phi$ values.}
\label{fig:precircuit_result}
\vspace{-0.15in}
\end{figure}

\textbf{Precircuit:}
To verify the functionality of the precircuit, we varied $V_{gs,in}$ from $1$ to $5~\rm{V}$, in increments of $0.1~\rm{V}$, for all four $\phi$ values. The reason not to start from $0~\rm{V}$ is that $V_{gs}$ should be greater than the threshold voltage, $V_{th} \approx 0.8~\rm{V}$, for the MOSFET to operate. The results, shown in Fig.~\ref{fig:precircuit_result}, are as expected for the case of $\phi=1,0.5~\rm{V}$. However, for the case of $\phi=0.25,0.125~\rm{V}$, the circuit maps to one level higher than expected for some voltages. For example, when $\phi=0.125~\rm{V}$, $V_{gs,in}=1.1~\rm{V}$ is mapped to $1.25~\rm{V}$ instead of $1.125~\rm{V}$. The reason for this may be the saturation effect of the Operational Amplifier~(OpAmp) used in the adder. Curated circuit design optimizations, outside the scope of this work, can help circumvent this limitation. 

\textbf{Encoding and Decoding:}
We used a $0.18~\mu\rm{m}$ technology n-channel MOSFET~(nMOS) with $W\cdot\mu\cdot C_{ox}/L = 155\times 10^{-6}~\rm{F/Vs}$, $V_{th} = 0.74~\rm{V}$, $\lambda = 0.037~\rm{V^{-1}}$ for evaluation purposes. $V_{ds}$ is varied from $4.5$ to $10~\rm{V}$, in increments of $0.1~\rm{V}$. The reason not to start from $0~\rm{V}$ is to ensure that the MOSFET is well into the saturation region. Discrete set of $V_{gs}$ values in the range $[1,5]~\rm{V}$ as per $\phi$ are considered, e.g., $V_{gs}=1,2,3,4,5~\rm{V}$ for $\phi=1~\rm{V}$; hence, for each $V_{gs}$, 55 values of $V_{ds}$ are considered. Upon applying these voltages to the MOSFET, the generated $I_{ds}$ values are recorded and sent to the digital receiver (no wireless channel), where the decoding process is done. At the receiver, each curve is processed independently and two consecutive $I_{ds}$ values from the same curve are used for decoding the correct $V_{gs}$ using the slope-matching technique. The results are shown in Fig.~\ref{fig:before_correction_delta_05} for $\phi=0.5~\rm{V}$, where the original values are shown using `+' and decoded values using `o'. We can see that some of the values are decoded incorrectly (where there are bare `+' without `o'). The reason is due to mismatch between two slopes---the slope calculated theoretically, $\lambda I_{ds}$, (varies with $V_{ds}$) is an approximation (i.e., valid only for $\lambda V_{ds} << 1$) of the actual slope calculated using the two-point formula (independent of $V_{ds}$).
To solve this problem, we used a range-checking technique where, if the decoded $V_{ds}$ value corresponding to the best (in terms of slope match) $V_{gs}$ value does not fall within the $V_{ds}$ range assumed at the transmitter $(4.5,10)~\rm{V}$, the next best $V_{gs}$ value (in terms of slope match) is chosen and the process is repeated iteratively. %
Using this correction logic, we are able to improve decoding accuracy.
To see the effect of $\phi$ on the decoding process, we varied it from $0.1$ to $1~\rm{V}$. Figure~\ref{fig:rmse_phi} shows the Root Mean Square Error~(RMSE) in $V_{gs}$ and $V_{ds}$ before and after the correction logic is applied. We have used separate axes for $V_{gs}$ and $V_{ds}$ as the error is higher in the case of $V_{ds}$. We can notice the following---(i)~before correction, the errors are high, up to about $0.3~\rm{V}$ for $V_{gs}$ and $7~\rm{V}$ for $V_{ds}$; (ii)~after correction, the error reduces up to about $0.1~\rm{V}$ for $V_{gs}$ and $2~\rm{V}$ for $V_{ds}$; (iii)~RMSE $\approx 0$ for $\phi \geq 0.4~\rm{V}$; increases steadily for $< 0.4~\rm{V}$.

\textbf{Variation with $\lambda$:}
Decoding accuracy varies with $\lambda$ (because of the reason mentioned above), which varies among different MOSFETs. To capture this effect, we varied $\lambda$ in the possible range $0.001$ to $0.2~\rm{V^{-1}}$ and plotted the RMSE of $V_{gs}$ and $V_{ds}$ (after correction logic is applied) by also varying $\phi$, as in Fig.~\ref{fig:rmse_lambda_phi_vgs_vds}. We observe that---(i)~for a given $\lambda$, lower $\phi$ performs worse; (ii)~the lowest possible $\lambda$ without noticeable degradation in RMSE keeps decreasing as $\phi$ is reduced; (iii)~for $\lambda < \approx 0.01$, the RMSE for all $\phi$ values has no noticeable degradation. Hence, it is desirable to consider those MOSFETs whose $\lambda<0.01$.

\textbf{Power Consumption:}
Our encoding design consists of precircuit and MOSFET. The power consumed by MOSFET is negligible compared to that of precircuit. Our precircuit primarily consists of OpAmps, comparators, multiplexers, and resistors, of which OpAmps are clearly the major contributors to the overall power consumption. The precircuit consumes one OpAmp for each stage and one for the final adder. For comparison purposes and to get an estimate of power consumption when our circuit is fabricated using the latest $nm$-Silicon technology, we use the same low power nano designs for the above components as considered in~\cite{Zhao16} ($8~\mu\rm{W}$ for OpAmp and $12.7~\rm{nW}$ for comparator). For 9 AJSCC levels ($\phi=0.5~\rm{V}$, 2-stages), the power consumption is $\approx 24~\mu\rm{W}$. On the other hand, Design~1~\cite{Zhao16} with $11$ levels consumes $130~\mu\rm{W}$ and Design~2~\cite{Zhao2018} with $8$ levels consumes $64~\mu\rm{W}$.

\textbf{Discussion:}
As pointed above, as $\phi<0.4~\rm{V}$, RMSE is non-zero for both $V_{gs}$ and $V_{ds}$. This suggests that having more than $10$ curves (i.e., AJSCC levels) in a single MOSFET will degrade the RMSE. To alleviate this undesired behavior, a multi-MOSFET architecture can be adopted. For example, in case $20$ AJSCC levels are desired, we can have four MOSFETs whose $V_{gs}$ values/curves are interwined so that there are only $5$ curves in each MOSFET, and $20$ combining all four. This achieves $\phi=0.2~\rm{V}$ without degradation in RMSE, unlike what we observe in Fig.~\ref{fig:rmse_phi}. All these four MOSFETs will need only 1 stage precircuit; and one precircuit can be reused for all four MOSFETs as only one of them is ON at a time. This reduces power consumption to $\approx 8~\mu\rm{W}$, making the circuit ultra low power. Additionally, it is possible that the performance of the MOSFET encoding varies with temperature~(T). To compensate for this undesired behavior, the above multi-MOSFET architecture can again be leveraged---e.g., consider two MOSFETs with opposing temperature sensitivities---so that the temperature sensitivity will be canceled in their combination. We will consider this as our future work.

\section{Conclusion and Future Work}\label{sec:conc}
We presented an ultra-low-power approach to realize analog joint source-channel coding in wireless transmissions and compress two sensor signals into one using a MOSFET device. We showed that the approach exhibits acceptable performance. As future work, we plan to evaluate the decoding behavior in the presence of different types of wireless channels.

\balance

\newpage

\bibliographystyle{IEEEtran} %
\bibliography{our_pubs,ref_ajscc_sensor,refs_biodegradable}

\end{document}